\documentclass{PoS}
\usepackage{amsmath}
\usepackage{placeins}

\title{Nucleon and $\Omega$ Baryon Masses with All-HISQ Fermions at the Physical Point}

\ShortTitle{Nucleon and $\Omega$ Baryon Masses with All-HISQ Fermions at the Physical Point}

\author{Ciaran~Hughes\\
        \fermilab\\
        E-mail:\email{chughes@fnal.gov}}

\author{\speaker{Yin~Lin}\\
        \chicago\\
        \fermilab\\
        E-mail: \email{yin01@uchicago.edu}}
        
\author{Aaron~S.~Meyer\\
        \bnl\\
        E-mail:\email{ameyer@quark.phy.bnl.gov}}

\newcommand{\fermilab}{Fermi National Accelerator Laboratory, Batavia, IL 60510, USA}

\newcommand{\chicago}{University of Chicago, Department of Physics, Chicago, IL 60637, USA}
\newcommand{\bnl}{Brookhaven National Laboratory, Upton, NY 11973, USA}

\abstract{We present the results of the nucleon and $\Omega$ baryon masses using staggered action for both valence and sea quarks. Three ensembles with the physical pion mass at approximate lattice spacings of $0.15$, $0.12$, and $0.088$fm are employed to extrapolate the masses to continuum and we obtain $M_N = 964(16)$ MeV and $M_\Omega = 1678(9)$. Both statistical and systematic uncertainties are included in the nucleon mass, whereas only the statistical uncertainty is accounted for in the $\Omega$ baryon mass.}

\FullConference{37th International Symposium on Lattice Field Theory - Lattice2019\\
		16-22 June 2019\\
		Wuhan, China}

\newcommand{\sun}[2]{\ensuremath{\text{SU}(#1)_{#2}}}
\newcommand{\gts}{\ensuremath{\text{GTS}}}

\begin{document}
\section{Introduction}
Lattice QCD calculations
with valence staggered fermions are complicated by 
remnant doublers, which 
are well understood in the meson sector \cite{Kilcup:1986dg, Golterman:1985dz} and have been successfully applied to many cutting-edge meson calculations (e.g., see \cite{Bazavov:2014pvz, Bazavov:2018omf, Chakraborty:2018iyb}). 
However, calculations of baryon properties with the staggered action have proven
to be challenging, complicated by the larger taste multiplicities than 
in the meson sector. Recent theoretical studies have been done to enumerate the staggered baryon operators~\cite{Golterman:1984dn, Bailey:2006zn,Gottlieb:2007ay}, but numerical simulations still remain unexplored. In these proceedings, we present our results of the nucleon and $\Omega$ baryon masses using the Highly Improved Staggered Quark (HISQ) action \cite{Follana:2006rc} for both valence and sea quarks. We use physical quark masses and 2+1+1 flavors of sea quarks. We focus on {\it i}) the nucleon mass as a precursor to a nucleon axial form factor calculation needed by the future long-baseline neutrino experiments~\cite{Alvarez-Ruso:2017oui}, and {\it ii}) the $\Omega$ baryon mass as an ideal observable to set the absolute lattice scale.

\section{Staggered Baryon Operators}
Staggered fermions have been successfully applied to computations of mesonic properties for many years, but calculations of baryonic quantities are mostly unexplored.
In Ref.~\cite{Golterman:1984dn}, Golterman and Smit described the procedures to construct baryonic interpolating operators with shifted staggered fields at rest, and Bailey~\cite{Bailey:2006zn} extended this idea to include \sun{3}{I} or \sun{2}{I} flavor symmetry and enumerated the expected ground states. We will summarize the methods described there, focusing on only the interpolating operators that we use for the nucleon and $\Omega$ baryon masses.

We are interested in classifying the continuum states that are irreducible representations (irreps) of 
\begin{equation}
    \sun{2}{S}\times \sun{8}{(I,T)} \supset \sun{2}{S}\times \sun{4}{T} \times \sun{2}{I} ,
\end{equation} where \sun{2}{S} is the spin symmetry, and $\sun{8}{(I,T)}\supset\sun{4}{T}\times\sun{2}{I}$ is the embedding group of \sun{4}{T} taste and \sun{2}{I} isospin. Parity and color quantum numbers are ignored for simplicity. The baryon wavefunctions must be totally anti-symmetric to obey Fermi-Dirac statistics, or totally symmetric under $\sun{2}{S}\times \sun{8}{(I,T)}$ since the color indices are anti-symmetrized. For the spin-$\frac{1}{2}$ particles we are interested in, the $\left(\frac{1}{2}, 168_M\right)$ irrep of $\sun{2}{S}\times \sun{8}{(I,T)}$ is the only irrep composed of three quarks that will satisfy the symmetry requirement. From this point forward, we will denote an irrep of a continuum group by its dimensions (with a bar on top if it is a conjugate representation) and a subscript, when necessary,  of $S$ (symmetric), $M$ (mixed), or $A$ (anti-symmetric). If the group is \sun{2}{}, however, we use the usual spin-$j$ fraction notation. In the above, $168_M$ is the $168$ dimensional irrep of \sun{8}{(I,T)} with mixed symmetry. We can then construct totally symmetric wavefunctions by combining $168_M$ with the mixed-symmetry spin-$\frac{1}{2}$ irrep.

To predict the mass of the $(\frac{1}{2}, 168_M)$ irrep, we can further decompose $168_M$ to irreps of $\sun{2}{I} \times \sun{4}{T}$ as
\begin{equation}
    168_M \rightarrow
    \left(\frac{3}{2}, 20_M\right)\oplus\left(\frac{1}{2}, 20_S\right)
    \oplus\left(\frac{1}{2},\bar{4}_A\right)
    \oplus\left(\frac{1}{2}, 20_M\right) .
\end{equation}
Baryons constructed from only a single quark taste should have the same masses as their QCD counterparts, and in this case, the nucleon mass. They belong to the $\left(\frac{1}{2}, 20_S\right)$ irrep that is symmetric in $\sun{4}{T}$. The isospin-taste symmetry then requires that all states in $168_M$ irrep to have the physical nucleon masses in the continuum limit regardless of isospin and/or taste quantum numbers. We will refer to these states as ``nucleon-like'' or ``$N$-like'' states to emphasize this fact. A similar nomenclature will be used for ``$\Delta$-like'' and ``$\Omega$-like'' states.

On lattice, the baryonic interpolating operators are constructed as irreps of $\sun{2}{I}\times \gts$, where \sun{2}{I} is again the isospin symmetry, and \gts~\cite{Golterman:1984dn}, the geometric timeslice group, is the discrete lattice subgroup of the spin, taste, and parity symmetries. There are only three possible fermionic irreps for \gts -- $8$, $8^\prime$, $16$, where the numbers again denote the dimensions. To access the nucleon properties, we will only use interpolating operators in the isospin-$\frac{3}{2}$, $16$ irrep. We can understand the particle content of this particular irrep following the group decomposition which yields one $N$-like state and three $\Delta$-like states in the lowest taste multiplet (i.e. only taste splittings and no orbital or radial excitations).

We can follow the same procedures for the $\Omega$ baryon, assuming an "isospin" symmetry for two species of strange quarks in the valence sector~\cite{Bailey:2006zn}. Conveniently, the "isospin"-$\frac{1}{2}$, $8^\prime$ irrep of $\sun{2}{I}\times \gts$ has only a single $\Omega$-like particle in the lowest taste multiplet, and we use this irrep to estimate its mass.

\section{Simulation Details}
We are using three $2+1+1$ HISQ \cite{Follana:2006rc} ensembles at the physical pion mass with approximate lattice spacings of $0.15$, $0.12$, and $0.088$ fm generated by the MILC collaboration \cite{Bazavov:2010ru}. We use the same HISQ quark action for the valence quarks. The $N$-like and $\Omega$-like baryon interpolating operators we use belong to the isospin-$\frac{3}{2}$, $16$ and ``isospin''-$\frac{1}{2}$, $8^\prime$ irreps of $\sun{2}{I}\times \gts$ respectively. There are four unique constructions, called classes, of the isospin-$\frac{3}{2}$, $16$ irrep made from three shifted staggered fermion fields. On the other hand, there is only one class of the ``isospin''-$\frac{1}{2}$, $8^\prime$ irrep for the $\Omega$-like baryon. All classes of operators are listed in \cite{Golterman:1984dn, Bailey:2006zn}. Additionally, we will show the results for the ``isospin''-$\frac{1}{2}$, 16 irrep to investigate taste-breaking effects on the $0.15$ fm ensemble in Sec.~\ref{sec:results}. 

There are 8(16) independent interpolating operators for the $8^\prime(16)$ irreps that can be used to construct two-point correlation functions. Fortunately, the correlation function is nonvanishing only when the same irrep components at the source and sink are paired together. We
average all of the available components to maximize the signal-to-noise ratio. We use the corner walls for the sources, and point-split local baryon operators at the sink~\cite{our_paper}.
The ensembles are fixed to Coulomb gauge to avoid putting gauge links at the source, reducing the number of propagator inversions; the gauge links are included at the sink, but we observe that the absence of 
sink links does not cause significant changes to our results.

\section{Results}
\label{sec:results}

For the $N$-like operators, we adopt both Bayesian \cite{Lepage:2001ym} and GEVP \cite{Blossier:2009kd} approaches to analyze our two-point correlation functions.
In the Bayesian analyses, we simultaneously fit to all $3\times 3 = 9$ correlators
(one operator class gives statistically imprecise results, so it is omitted from the analysis)
to multi-exponential functions with appropriate priors on masses. In the GEVP approach, we solve 
the equation
\begin{align}
	\frac{1}{4} \bigg\{ &[\mathbf{C}(t_0-1)]^{-1} \mathbf{C}(t-1)  + 2[\mathbf{C}(t_0)]^{-1} \mathbf{C}(t) 
	+ [\mathbf{C}(t_0+1)]^{-1} \mathbf{C}(t+1) \bigg\} v_1(t) = \lambda_1(t, t_0) v(t)
	\label{eq:gevp}
\end{align}
where $\mathbf{C}(t)$ is the 
($3\times3$)
matrix of correlators, and $\lambda_1$ and $v_1$ are the eigenvalue and eigenvector that
correspond to the nucleon-like ground state. In these proceedings, we 
show results only from the GEVP analyses as the results from our Bayesian analyses agree within uncertainties. 

\begin{figure*}[!htbp]
\begin{center}
\begin{tabular}{cc}
 \includegraphics[scale=0.37]{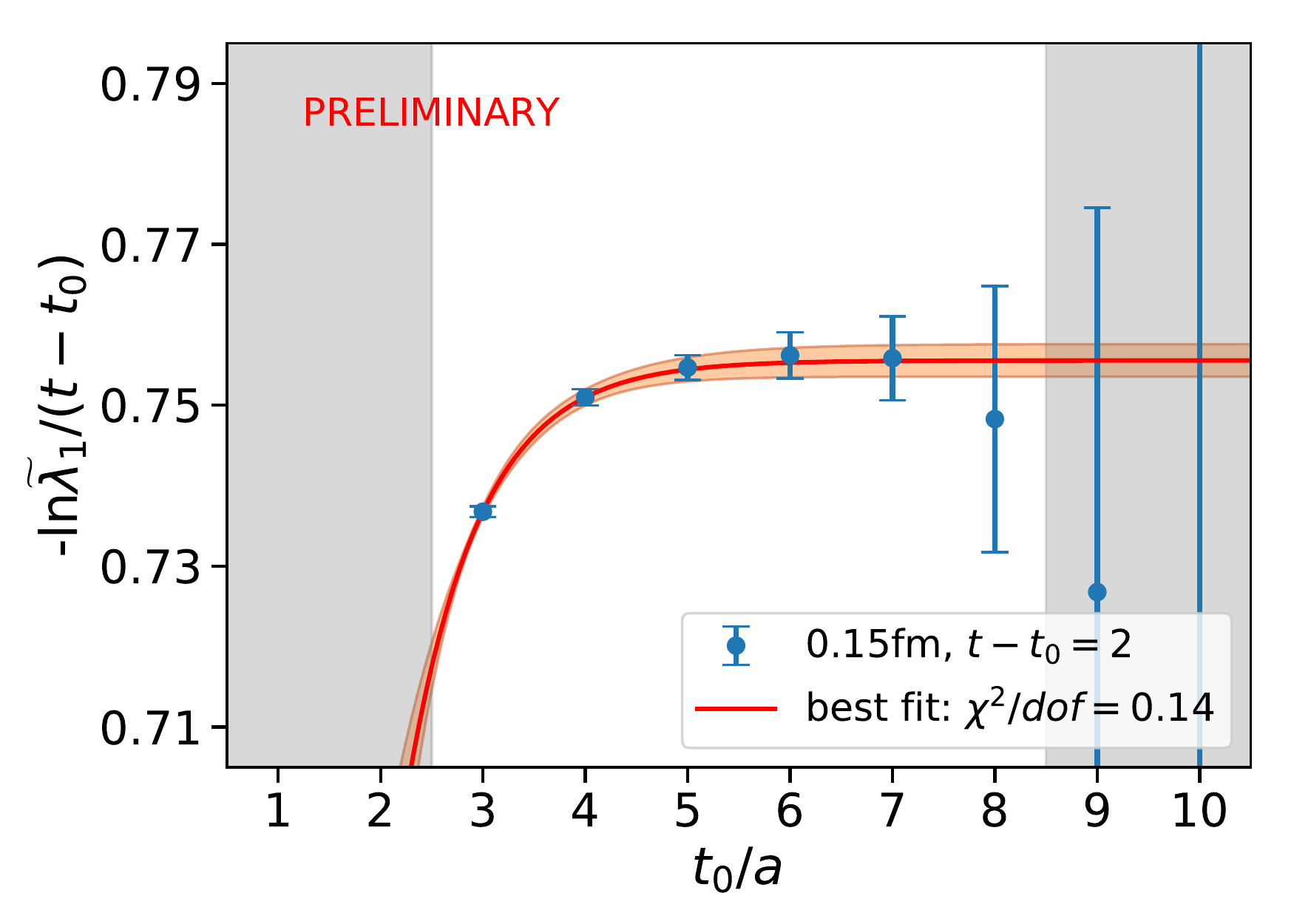}&
 \includegraphics[scale=0.37]{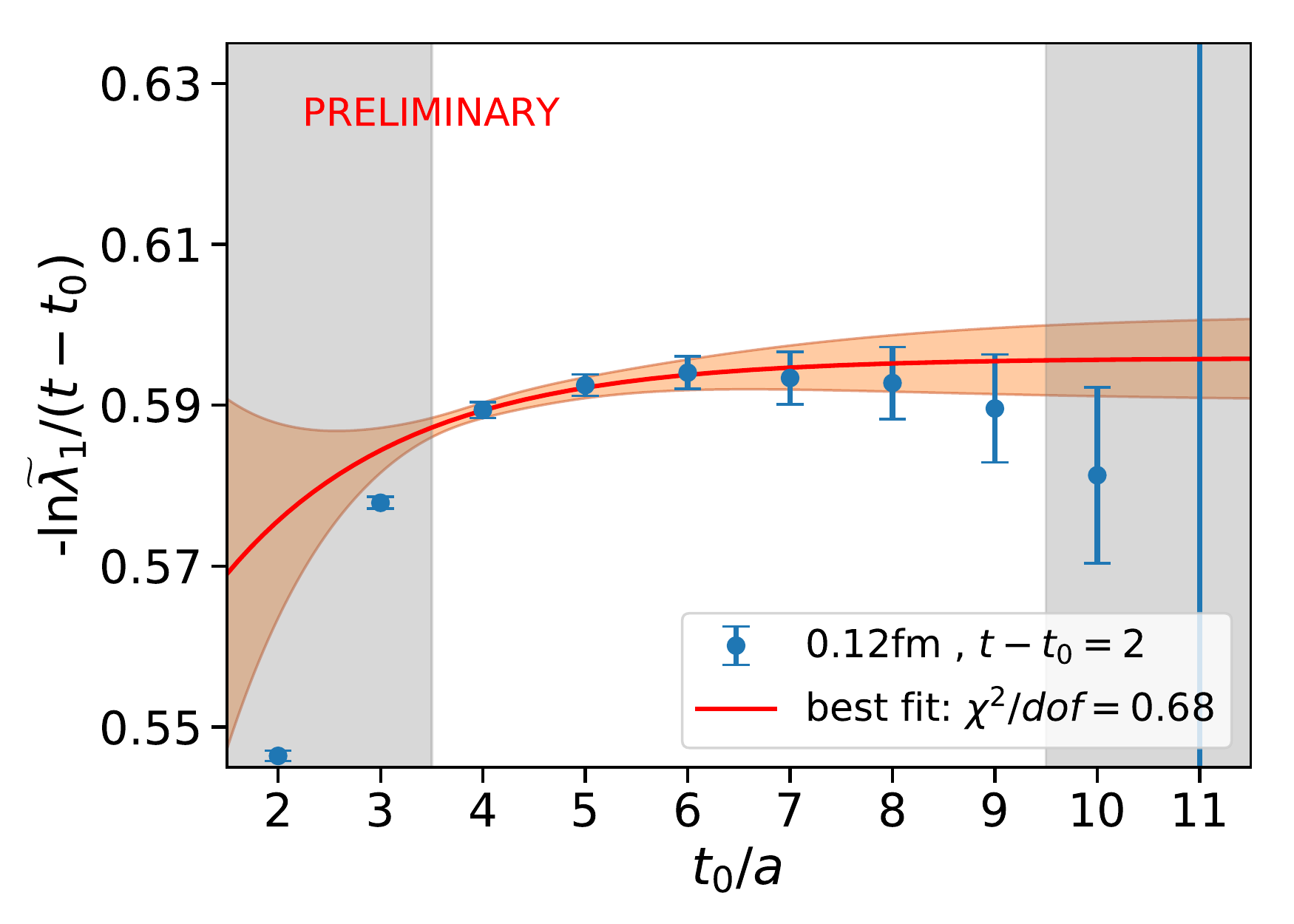}\\
 \multicolumn{2}{l}{\centerline{\includegraphics[scale=0.37]{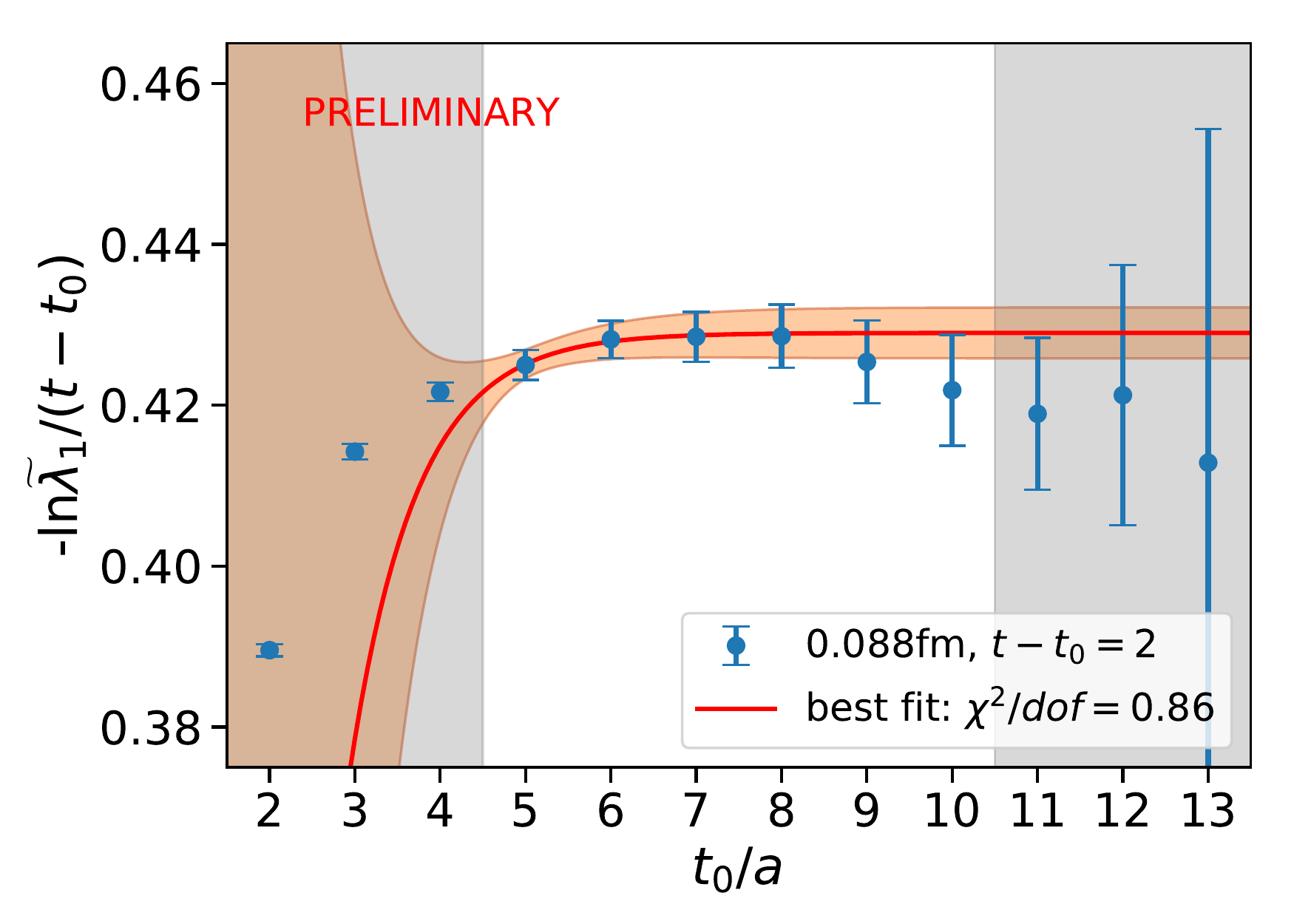}}}
\end{tabular}
\caption{Effective masses of the eigenvalues for the $N$-like baryon 
for the $a\approx 0.15$ fm ensemble (top left), $a\approx 0.12$ fm ensemble (top right), and $a\approx 0.088$ fm ensemble (bottom).
The data from solving Eq.~(\ref{eq:gevp}) are shown as blue circles, the white regions are the fitting ranges, and the red lines (bands) are the best fit curves (errors) from the plateau fit with an exponential term.}
\label{fig:nucleon}
\end{center}
\end{figure*}

\begin{figure*}[!htbp]
\begin{center}
\begin{tabular}{cc}
 \includegraphics[scale=0.38]{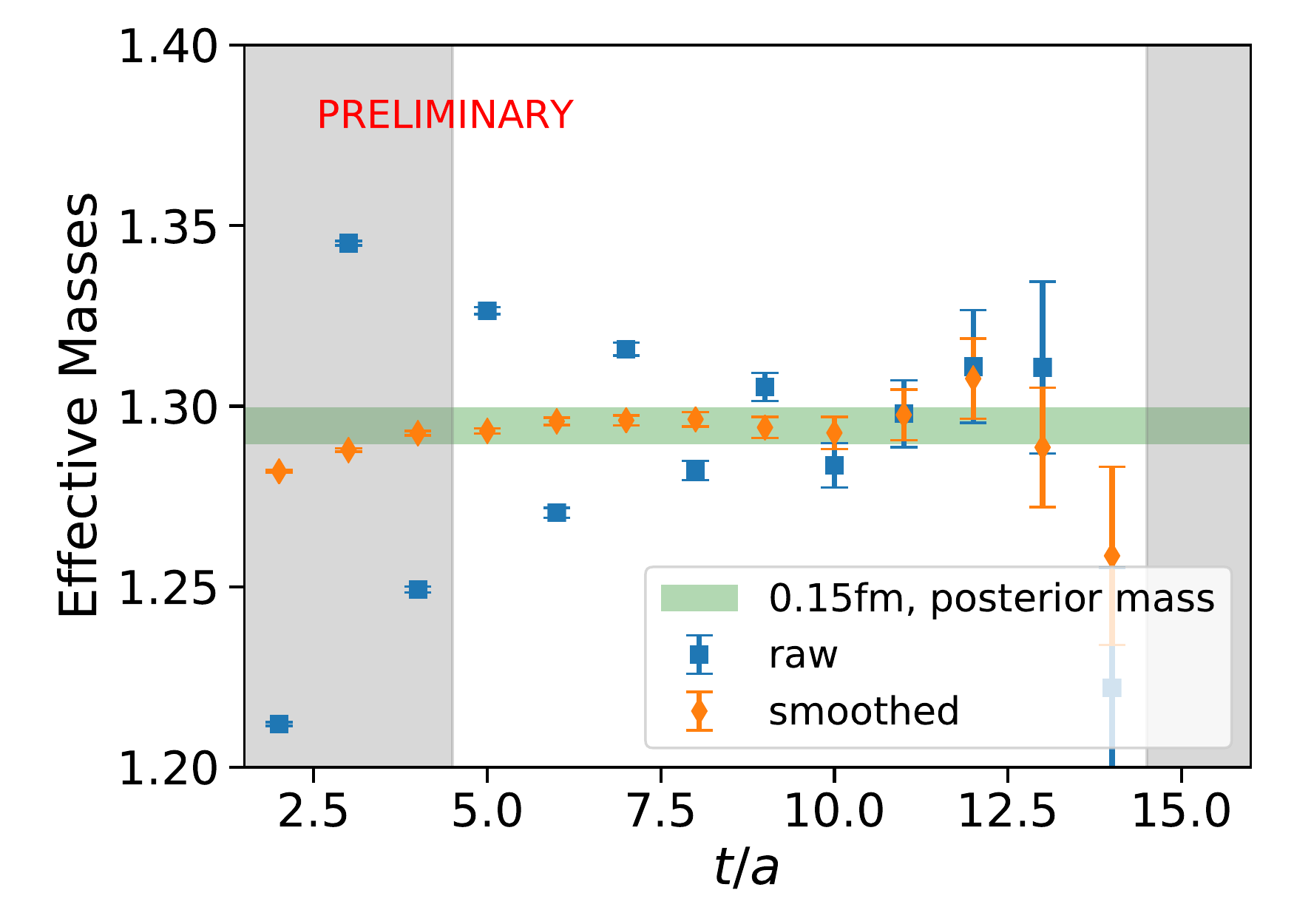}&
 \includegraphics[scale=0.38]{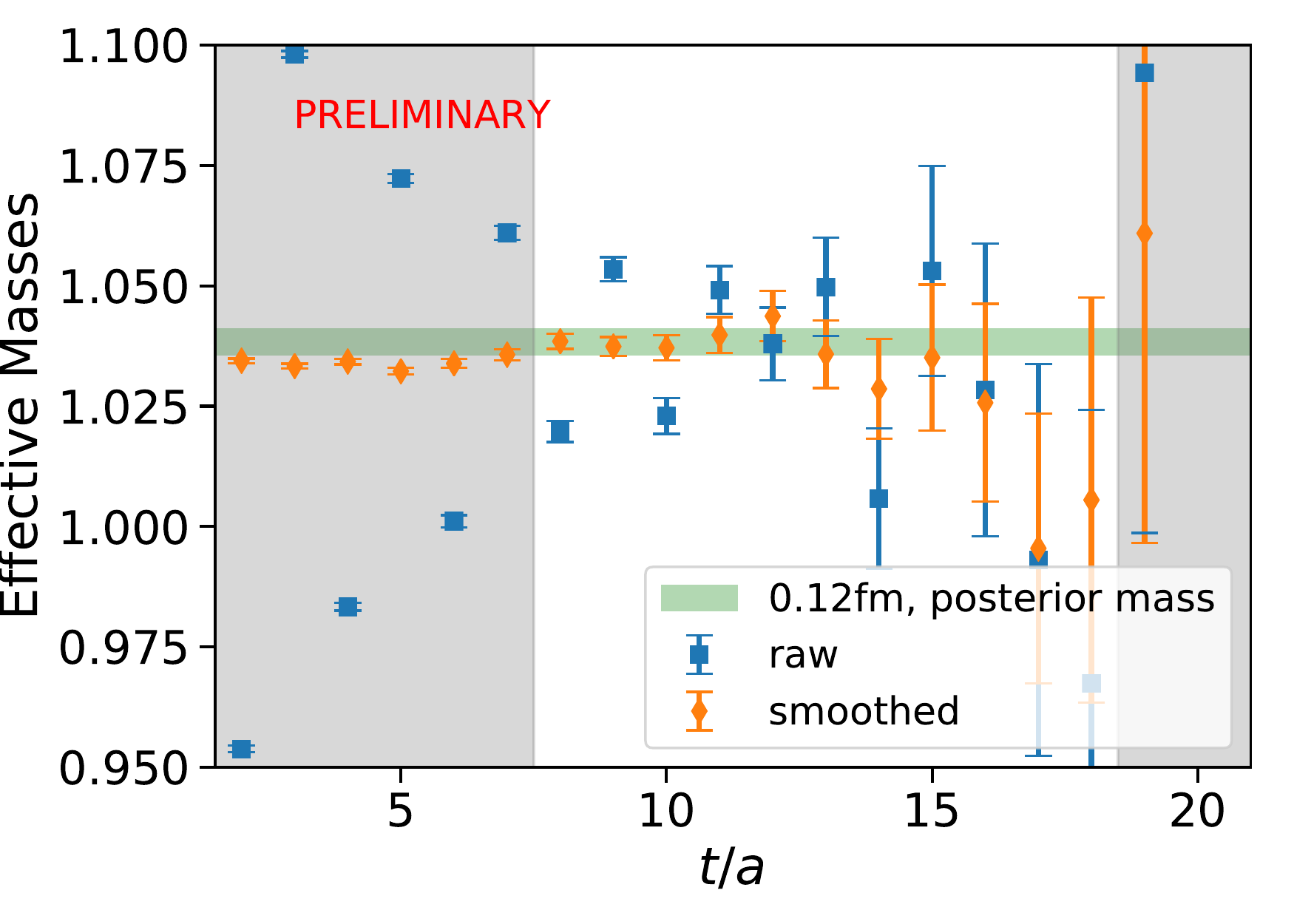}\\
 \multicolumn{2}{l}{\centerline{\includegraphics[scale=0.4]{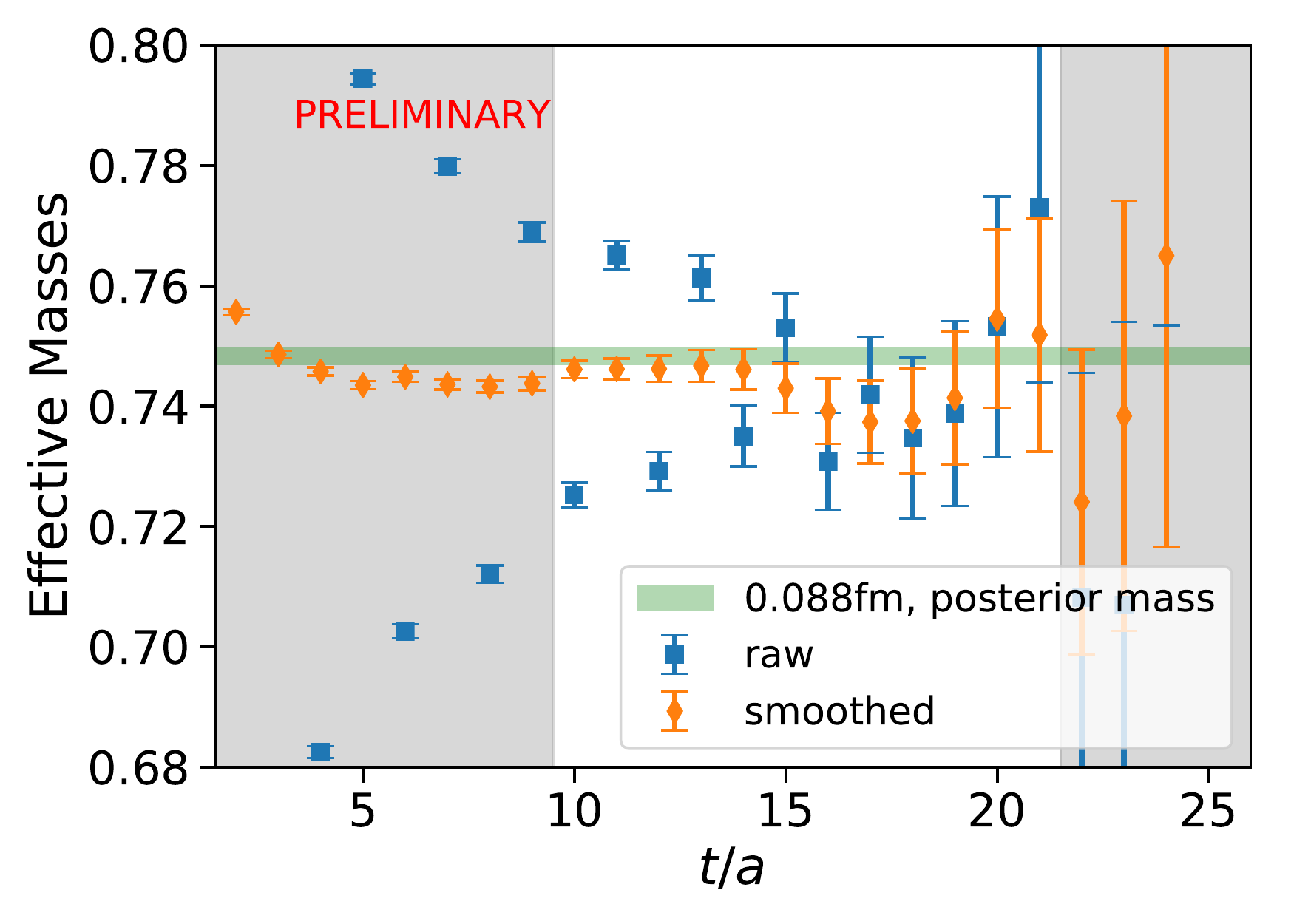}}}
\end{tabular}
\caption{Effective masses for the $\Omega$-like baryon across three lattice spacings. The white regions are the fit ranges, and the green bands are the posterior masses from multi-exponential Bayesian fits.
The blue squares are the traditional effective mass (Eq.~(\ref{eq:meff})), and the orange circles are the smoothed effective mass(Eq.~(\ref{eq:meff_smooth})).
Note that we use a slightly different smearing radius for $a\approx 0.088$ fm ensemble, so the excited state contamination is different.
}
\label{fig:omega}
\end{center}
\end{figure*}

The analyses for the $\Omega$-like baryon are straightforward with a single operator so we fit the correlators directly to the multi-exponential functions. To visualize the $\Omega$-like baryon correlators, in addition to the usual effective masses, $M_{\text{eff}}(t)$, we also plot the smoothed effective masses \cite{DeTar:2014gla}, $\overline{M}_{\text{eff}}(t)$, which are designed to alleviate the staggered oscillation. They are defined as
\begin{align}
    aM_{\text{eff}}(t) &\equiv \frac{1}{\tau}\left(\frac{C(t)}{C(t+\tau)}\right)\label{eq:meff}\\
    a\overline{M}_{\text{eff}}(t) &\equiv \frac{1}{4}\bigg(2M_{\text{eff}}(t) + M_{\text{eff}}(t-1) + M_{\text{eff}}(t+1)\bigg)\label{eq:meff_smooth}
\end{align}
where $\tau=2$. The results are plotted in Figures \ref{fig:nucleon} and \ref{fig:omega}.

\begin{figure*}[!htbp]
\begin{center}
\begin{tabular}{cc}
\includegraphics[scale=0.4]{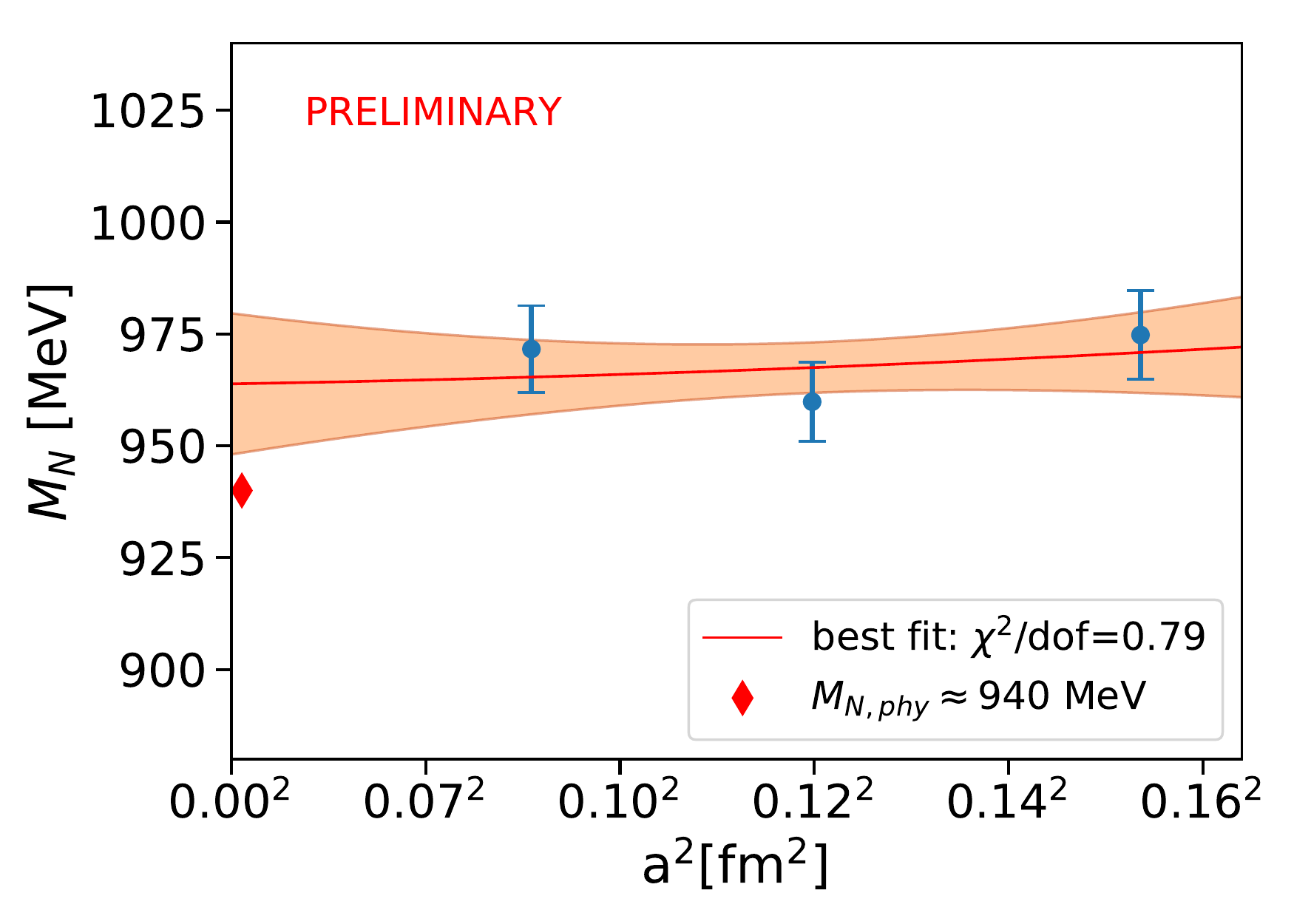}
\includegraphics[scale=0.4]{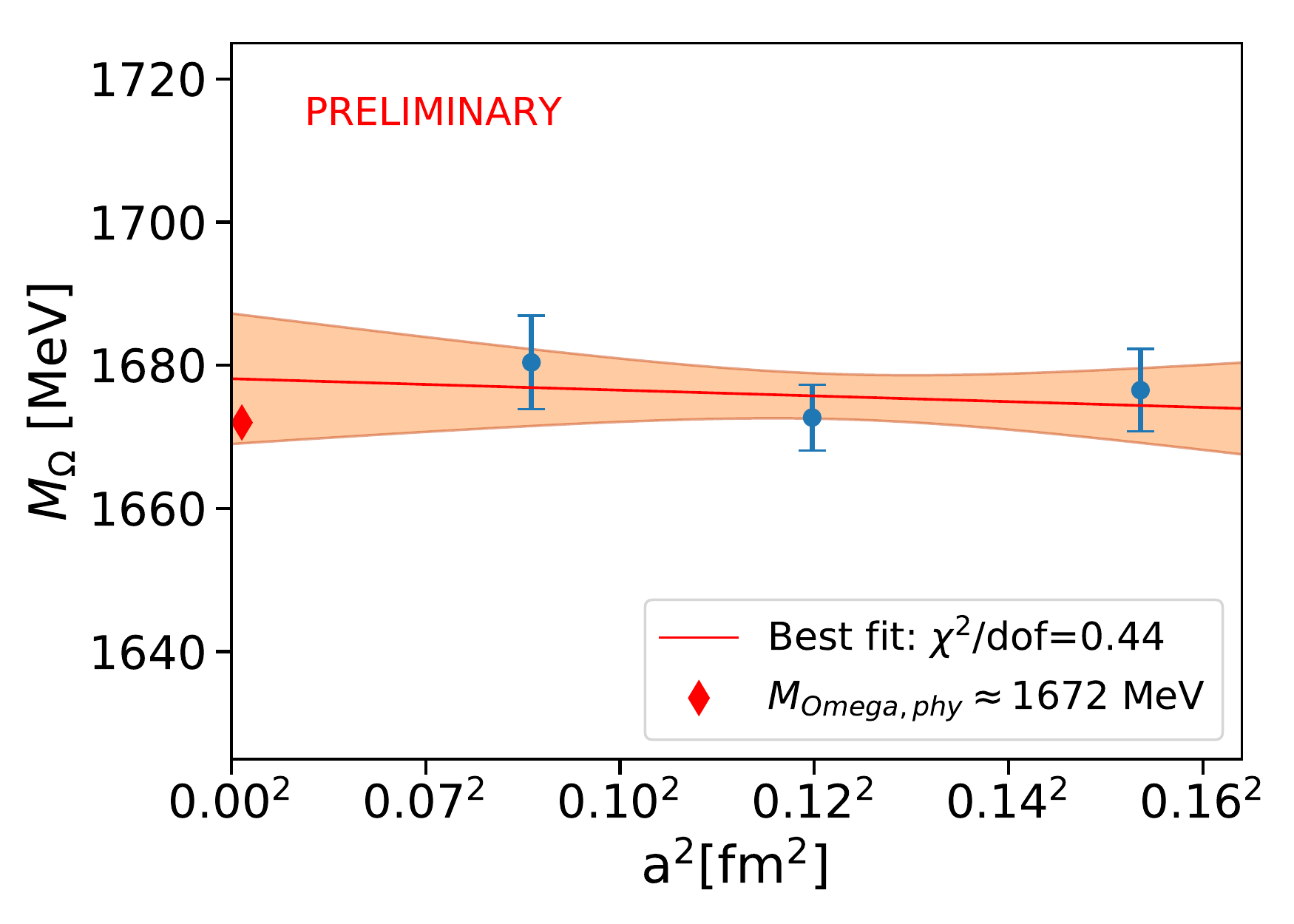}
\end{tabular}
\caption{Continuum extrapolations for the $N$-like (left) and $\Omega$-like (right) masses using equation (\ref{eq:con}).}
\label{fig:continuum}
\end{center}
\end{figure*}

With three lattice spacings, we can extrapolate the posterior masses to the continuum by using the fit function
\begin{equation}
    M_{\text{lat.}} = M_{\text{cont.}}\bigg(1+o_2(\Lambda_{\text{QCD}}a)^2 + o_4(\Lambda_{\text{QCD}}a)^4\bigg) ,
    \label{eq:con}
\end{equation}
where we use the lattice spacings in the mass-independent $f_{p4s}$ scheme defined in \cite{Bazavov:2015yea} and $\Lambda_{\text{QCD}} = 500$ MeV. We perform the Bayesian fits with unconstrained prior on $o_2$ and $0(1)$ prior on $o_4$. The results are shown in Figure \ref{fig:continuum}.

The continuum masses for the nucleon and $\Omega$ baryon we obtain are $M_N = 964(16)$ MeV and $M_\Omega = 1678(9)$ MeV respectively. We have listed the full error budget for the nucleon mass in \cite{our_paper} with both statistical and systematic errors accounted for. The $\Omega$ baryon mass error is statistical only.

\begin{figure*}[!htbp]
\begin{center}
\includegraphics[scale=0.43]{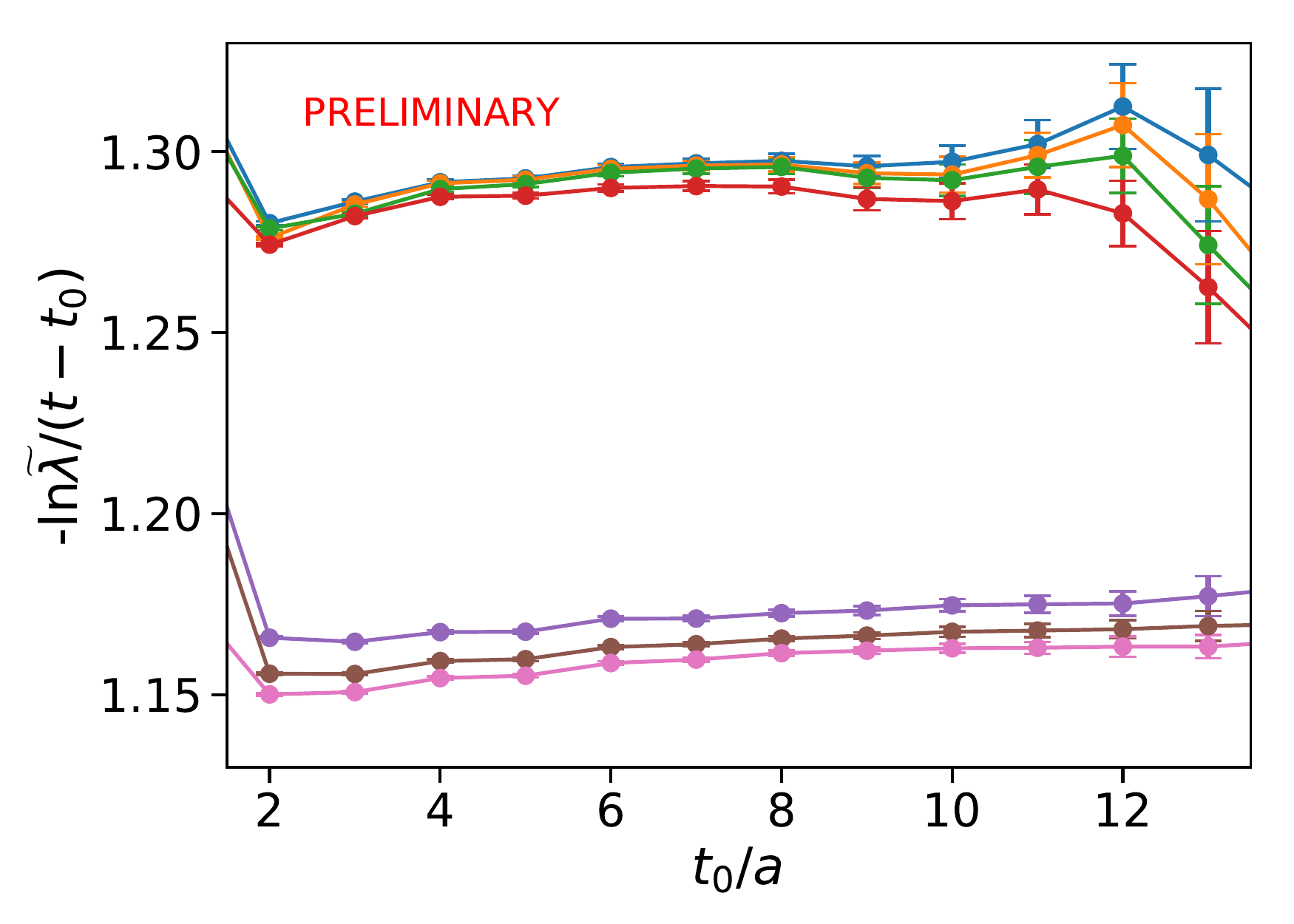}
\caption{GEVP analysis of the "isospin"-$\frac{1}{2}$, 16 irrep operators built from strange valence quarks on the $a\approx 0.15$ fm ensemble. Different colored lines correspond to
different eigenvalues obtained by solving equation (\ref{eq:gevp}) with $t-t_0 = 2$.}
\label{fig:taste}
\end{center}
\end{figure*}

So far, we have only shown the correlator results with a single ground state using isospin-$\frac{3}{2}$, $16$ irrep operators for the $N$-like and "isospin"-$\frac{1}{2}$, $8^\prime$ irrep for the $\Omega$-like states. In general, however, a staggered baryon operator overlap with
multiple states that become
degenerate in the continuum limit. It is interesting theoretically to investigate how well we can separate these different taste states and estimate the size of taste breaking effects that
occur on a finite lattice. With these goals in mind, we have plotted
the GEVP results of the "isospin"-$\frac{1}{2}$, $16$ irrep operator in Figure \ref{fig:taste} with strange valence quarks and $0.15$ fm ensemble. There are seven classes for this particular irrep which interpolate to three $N_s$ (equivalent to the nucleon, but made of two degenerate species of strange valence quarks) and four $\Omega$-like states. We find that the effective masses for the $\Omega$-like states match with
the expected spectrum, and we can
disentangle different baryon tastes
despite the smallness of the mass separations. The findings are promising and warrant more systematic studies to understand the staggered baryon taste splittings.

\section{Conclusions}
In these proceedings,
we demonstrate the methods to calculate the nucleon and $\Omega$ baryon masses using HISQ quarks for both valence and sea sectors. With the cost advantage of staggered fermions,
we are able to perform all simulations on the physical pion mass ensembles at the approximate lattice spacings of $a\approx 0.15$, $0.12$, and $0.088$ fm. The continuum-extrapolated
masses for the nucleon and $\Omega$ baryons are $M_N = 964(16)$ MeV and $M_\Omega = 1678(9)$, where the nucleon mass error comprises both statistical and systematic components \cite{our_paper} and the $\Omega$ mass error is statistical only. The nucleon mass is found to be high, which we attribute to the statistical fluctuation
 on the $a\approx 0.088$ fm ensemble. Furthermore, we demonstrate our control of the nearly-degenerate
baryon tastes with a large basis of operators. Our results for the nucleon mass mark
the first step towards a calculation of nucleon matrix elements with the staggered action, and 
the $\Omega$ baryon mass result is immediately applicable to
achieving a more accurate lattice scale determination.

\end{document}